\documentclass[aps,prl,twocolumn,showpacs]{revtex4}
\usepackage{bm}
\usepackage{graphicx}

\renewcommand{\vec}[1]{\bm{#1}}

\begin{document}
\title{Bulk viscosity and the conformal anomaly in the pion gas}
\author{D. Fern\'andez-Fraile}
\email{danfer@fis.ucm.es}
\author{A. G\'omez Nicola}
\email{gomez@fis.ucm.es} \affiliation{Departamento de F\'isica
Te\'orica II, Universidad Complutense, 28040 Madrid, Spain.}
\date{\today}

\begin{abstract}
We calculate  the bulk viscosity of the massive pion gas within
Unitarized Chiral Perturbation Theory.  We obtain a low temperature
peak arising from explicit conformal breaking due to the pion mass
and another peak near the critical temperature, dominated by the
conformal anomaly through gluon condensate terms. The correlation
between bulk viscosity and conformal breaking supports a recent QCD
proposal. We discuss the role of resonances, heavier states and
large-$N_c$ counting.
\end{abstract}

\pacs{11.10.Wx, 12.39.Fe, 25.75.-q}
\maketitle

The matter produced after thermalization in relativistic heavy ion
collisions  behaves nearly as a perfect fluid \cite{Adams:2003am}.
Deviations  are seen mainly in elliptic flow and can be reasonably
explained with a small shear viscosity over entropy density ratio
$\eta/s<0.5$ \cite{viscotheor}, whereas bulk viscosity $\zeta$  is
generically assumed to be
 negligible. However,  it has been recently proposed
 \cite{Karsch:07}  that $\zeta$  might be
large near the QCD phase transition. If $\zeta/s$ is comparable to
$\eta/s$ near the critical point (where indeed the latter is
expected to have a minimum) interesting physical possibilities arise
such as radial flow suppression, modifications of the hadronization
mechanism \cite{Karsch:07} or clustering at freeze-out
\cite{Torrieri:2007fb}. The argument of \cite{Karsch:07} is that,
following the QCD sum rules  in \cite{Shushpanov:1998ce}, one can
relate $\zeta$ with the trace anomaly:
\begin{equation}
\label{kharzeev}
   \zeta (T) =\frac{1}{9\omega_0(T)}\left[T^5\frac{\partial}{\partial
    T}\frac{\langle\theta\rangle_T-\langle\theta\rangle_0}{T^4}+16\vert\epsilon_0\vert\right],
\end{equation}
with
$\langle\theta\rangle_T\equiv\langle T^\mu_\mu\rangle_T=
\epsilon-3P$,  $T^\mu_\nu$ the energy-momentum
tensor, $\epsilon$ the energy density, $P$ the pressure and
$\epsilon_0=\langle\theta\rangle_0/4$ in  vacuum. To derive
(\ref{kharzeev}), a particular ansatz has been used for
$\rho(\omega)$, the $\langle\theta\theta\rangle$ spectral function
at zero spatial momentum, with $(\rho/\omega)(0)=9\zeta/\pi$ and
$9\zeta\omega_0=2\int_0^\infty (\rho/\omega)d\omega$. Eq.
(\ref{kharzeev}) implies then a large bulk viscosity near the QCD
transition, from the $\langle\theta\rangle_T$ peak  observed in
the lattice \cite{Cheng:08}, more or less pronounced depending on
the transition order \cite{Karsch:07}. However, this argument has
been recently criticized on the basis of the $\int_0^\infty
(\rho/\omega)$ convergence   and
 parametric dependence with the QCD coupling constant
\cite{Moore:2008ws}. On the other hand, estimates of $\zeta$ from
lattice data show that $\omega\delta(\omega)$ terms and
large-$\omega$ non-thermal contributions have to be properly
accounted for in  spectral functions
\cite{Meyer:2007dy,Huebner:2008as}.

\begin{figure}
\centerline{\includegraphics[width=4.5cm]{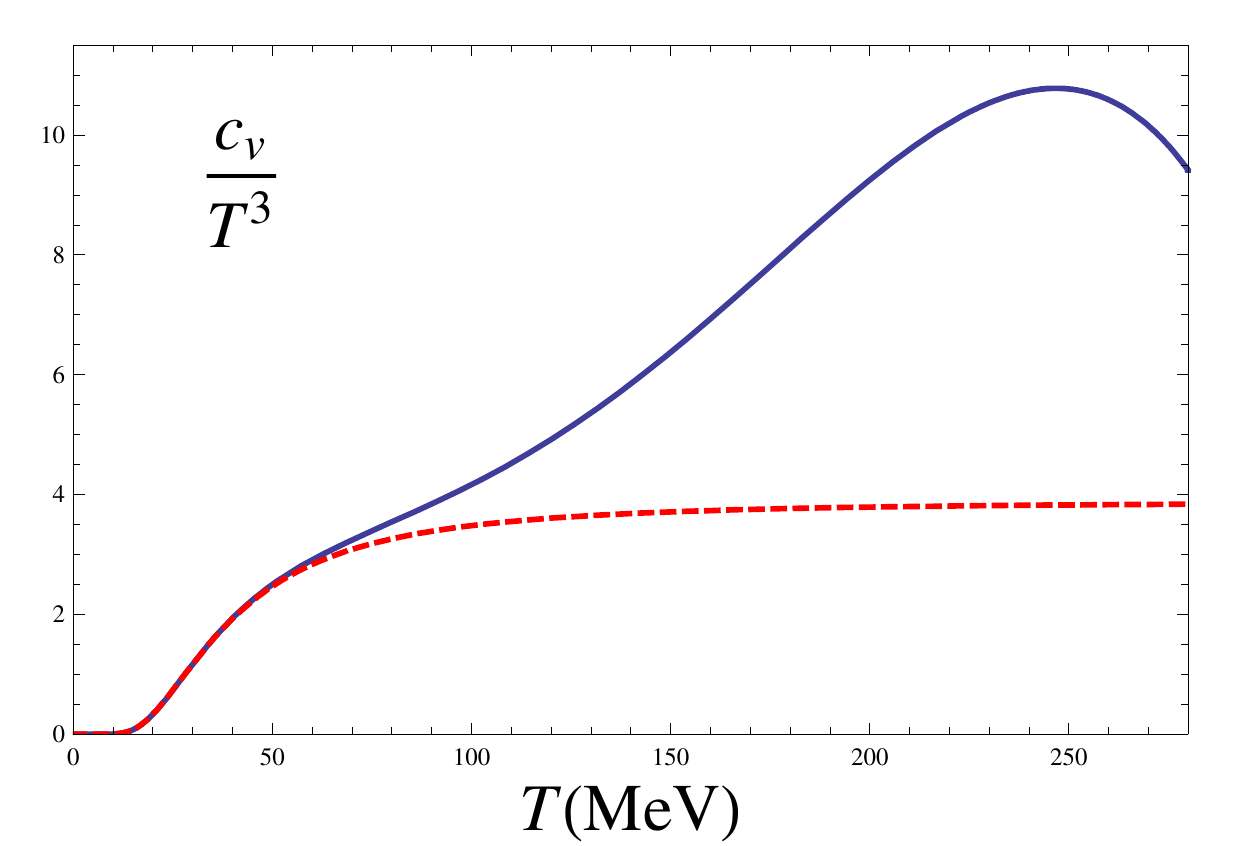}\includegraphics[width=4.5cm]{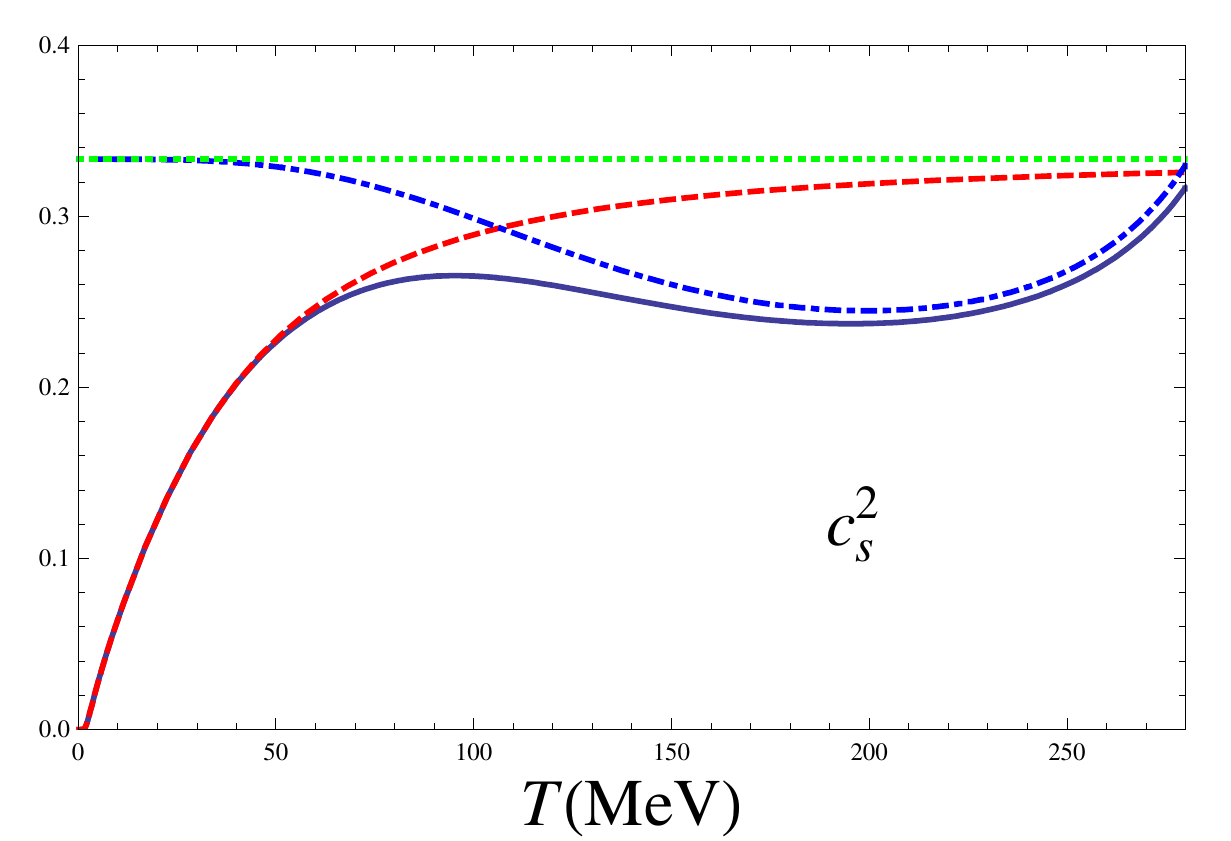}}
\caption{Specific heat (left) and speed of sound squared (right) for
the pion gas. The red dashed line is the $\mathcal{O}(T^6)$
calculation, and the continuous blue line  the $\mathcal{O}(T^8)$
one. The green dotted line is the ultra-relativistic limit
$c_s^2=1/3$. The dashed-dotted blue line is the chiral limit result
to $\mathcal{O}(T^8)$. }\label{figthermod}
\end{figure}

It is therefore of great importance to study QCD regimes where one
can rely on analytic calculations, in order to clarify the validity
of the above  proposal without appealing directly to lattice data.
In the weak coupling regime, valid for very high temperatures,
$\zeta/\eta$ has been found to be parametrically small
\cite{Arnold:2006fz}. Another
 regime where one can perform analytic calculations is
low-energy QCD, where the system consists primarily of a meson gas
and, for low temperatures, one can rely on Chiral Perturbation
Theory (ChPT) \cite{Gerber:89}. In this regime, we have recently
shown \cite{Fernandez-Fraile:05,Fernandez-Fraile:07}, within
Linear Response Theory (Kubo's formulae), that the usual
ChPT power counting must be extended to account for $1/\mathnormal{\Gamma}_p$
contributions arising in transport coefficients. Here, $\mathnormal{\Gamma}_p$ is the thermal width
 of a pion  with three-momentum $\vec{p}$,  in which the $\pi\pi$ total elastic cross section
enters linearly in the dilute gas regime \cite{width}. Performing
the power counting, which includes a detailed analysis of
ladder-type diagrams considered in \cite{Jeon:1994if}, the
leading-order ChPT contribution comes from a one-loop meson
diagram with $\mathnormal{\Gamma}_p\neq 0$ internal lines. An
essential point is to include unitarity corrections in
$\mathnormal{\Gamma}_p$ to describe correctly the temperature
behavior  as the system approaches chiral restoration. We neglect
inelastic $2\pi\leftrightarrow 4\pi$ reactions restoring particle
number equilibrium, which  are suppressed in  our counting and
yield chemical relaxation times about ten times larger than the
 plasma lifetime \cite{Song:1996ik}.  Thus, our bulk viscosity is meaningful for
 the pion gas formed in heavy ion collisions, which conserves approximately pion number between
chemical and thermal freeze-out, as confirmed by particle spectra data analyses with a pion chemical potential  \cite{Hung-Kolb}.  If $\zeta$ is defined in complete chemical
equilibrium, then particle-changing processes dominate
\cite{Jeon:1994if}. The dominance of elastic processes for $\zeta$
in the pion gas holds also in kinetic theory
\cite{Prakash:1993bt,Chen:07,Gavin:1985ph}. With our approach we
have also obtained  $\eta/s$ developing a minimum compatible
 with AdS/CFT bounds   with values in good agreement with
 kinetic theory  \cite{Prakash:1993bt,Dobado:2006hw} and phenomenological estimates on elliptic flow. This is the theoretical basis of the present work, where we
will
 analyze within  ChPT  the correlation between bulk viscosity and the conformal anomaly  in
the pion gas regime, studying the origin of the different
contributions to conformal breaking for physical massive pions.
Thus, we start with Kubo's formula:
\begin{equation}
\zeta (T)=\frac{1}{2}\lim_{\omega\rightarrow
0^+}\frac{\partial}{\partial \omega}\int\mathrm{d}^4x\
\mathrm{e}^{\mathrm{i}\omega x^0}\
\langle[\hat{\mathcal{P}}(x),\hat{\mathcal{P}}(0)]\rangle,
\label{kubo}
\end{equation}
where the modified pressure operator
$\hat{\mathcal{P}}\equiv-T^i_i/3-c_s^2T_{00}$,  the squared
speed of sound $c_s^2=\partial P/\partial\epsilon=s/c_v$,
$s=\partial P/\partial T$ and the specific heat $c_v=\partial
\epsilon/\partial T=T\partial s/\partial T$. We follow the
conventions of \cite{Hosoya:1983id}, where $\zeta$ is defined as
the change in the pressure produced by a gradient in the flow
velocity, relative to  equilibrium. This leads to the
 correlator in (\ref{kubo}), which is the adequate
one to be used within perturbation theory
\cite{Arnold:2006fz,Moore:2008ws}. In
lattice analyses,  one works with the Lorentz invariant $\theta$
instead. In our approach these two correlators are not equivalent,
since
  the leading order in $1/\mathnormal{\Gamma}_p$ for perturbative $T_{00}$ commutators does not vanish for zero spatial momentum. As
  we shall see, sticking to the original definition (\ref{kubo}) leads naturally to the expected conformal properties and asymptotic behavior of the bulk viscosity. Following \cite{Fernandez-Fraile:05}, we calculate then the  spectral
function ($\hat{\mathcal{P}}$ commutator) in (\ref{kubo}) in the
imaginary-time formalism, picking up the dominant contribution in
$1/\mathnormal{\Gamma}_p$ (pinching pole) of the analytically
continued retarded correlator.   That
term is purely imaginary and gives the dominant effect in
the spectral function at zero momentum and small energy.  Thus, to leading order: \begin{equation}\label{bulk}
\zeta (T)=\int\limits_0^\infty\mathrm{d}p
 \frac{3p^2(p^2/3-c_s^2E_p^2)^2}{4\pi^2 T
E_p^2\mathnormal{\Gamma}_p} n_\mathrm{B}(E_p)[1+n_\mathrm{B}(E_p)],
\end{equation}
with $n_\mathrm{B}(x)\equiv 1/(\exp(x/T)-1)$  the Bose-Einstein
distribution function,  $E_p\equiv\sqrt{p^2+M_\pi^2}$ and where
the leading $\mathcal{O}(p^2)$ order in $T^{\mu\nu}$ has been
retained in the vertex.  Now, we get $c_s^2$ in (\ref{bulk}) from
$P$ calculated up to $\mathcal{O}(T^8)$ in \cite{Gerber:89}.  In
Fig.\ref{figthermod} we see that to $\mathcal{O}(T^6)$, both  the
specific heat and the speed of sound increase monotonically,
$c_s^2$ approaching the ultra-relativistic limit of $1/3$
corresponding to a gas of free massless pions. To that
order, since the distribution function is peaked around $p\sim T$  for $T\gg m_\pi$, we see that (\ref{bulk}) vanishes asymptotically for large
temperatures,  as expected for
conformally invariant systems
\cite{Fernandez-Fraile:07,Arnold:2006fz,Hosoya:1983id,Gavin:1985ph,Prakash:1993bt}. In
fact, from (\ref{bulk}) we get for massless pions (chiral limit)
$\zeta=15(1/3-c_s^2)^2\eta$, consistently with
\cite{Horsley:1985dz} and parametrically with high-$T$ QCD
\cite{Arnold:2006fz}. The crucial point here is that taking one
more order in the pressure $c_v$ grows, reaching a maximum at
about $T_c\simeq $ 220 MeV. The speed of sound attains then a
minimum at $T_c$ which will alter the
 behavior of $\zeta (T)$. This is the critical behavior of a
$O(4)$-like crossover, as expected for two massive flavors at zero
chemical potential. A physical interpretation is that, although
temperature tends to erase mass scales, chiral interactions are
enhanced and produce in the critical region a significant,
nonperturbative,  conformal breaking reflected in $c_s^2\neq 1/3$.
Note that, although in the massive case
 $T_c$ is near the chiral restoration temperature $T_c^\chi$
where the order parameter $\langle \bar q q\rangle_T$ vanishes
\cite{Gerber:89}, in the chiral limit $T_c^\chi\simeq$ 170 MeV,
while $T_c$ is almost unchanged.

\begin{figure}
\centerline{\includegraphics[width=6.5cm]{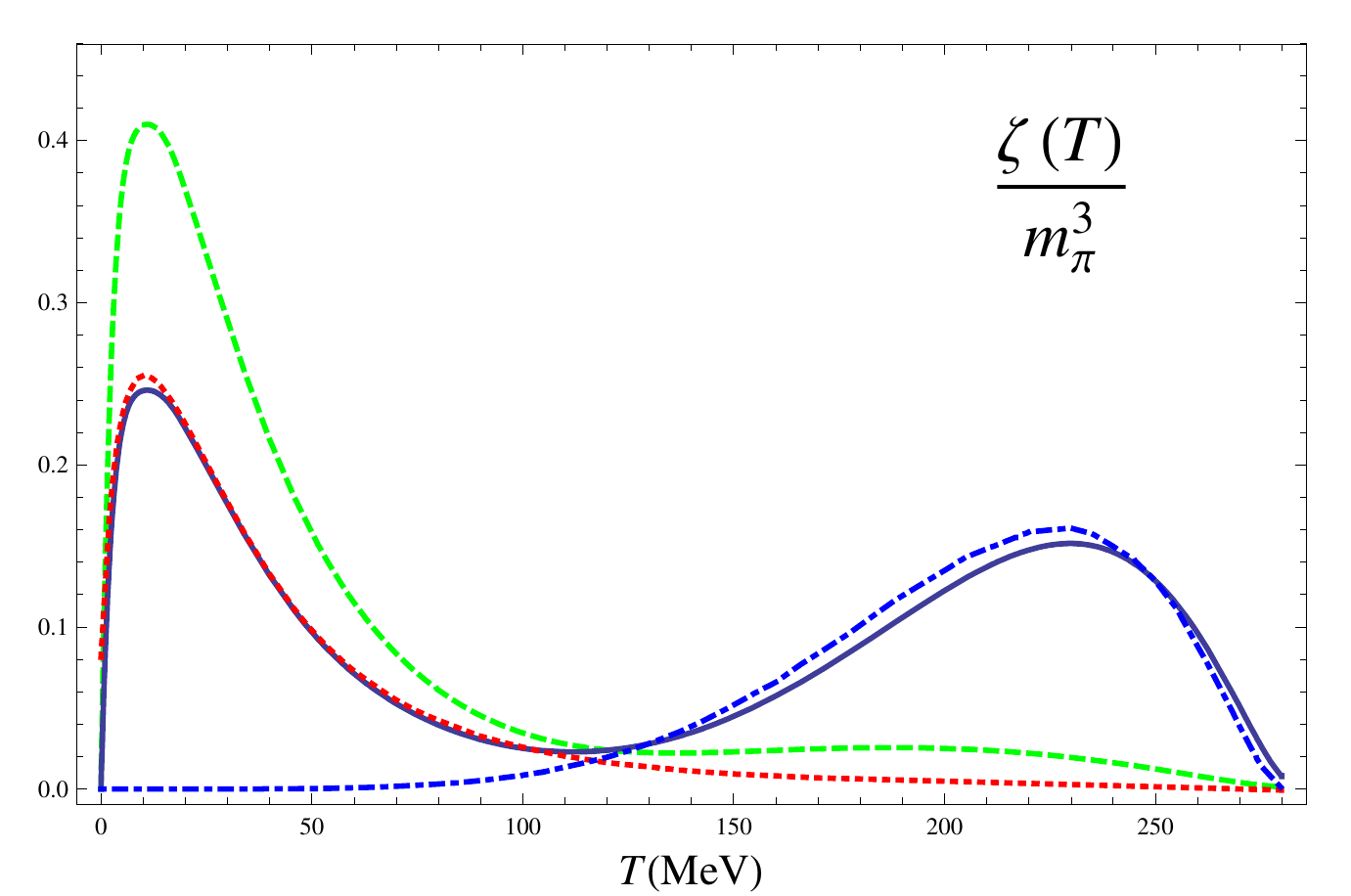}} \caption{Bulk
viscosity of the pion gas. The full blue line is the unitarized
result with $c_s^2$  to $\mathcal{O}(T^8)$ and the dashed-dotted
blue one is the same calculation in the chiral limit. The dashed
green line is the nonunitarized result at the same order. The dotted
red line is unitarized with $c_s^2$  to
$\mathcal{O}(T^6)$ and lies very close to the nonunitarized curve, which is not displayed.}\label{figbulk}
\end{figure}

We plot our result for the bulk viscosity in Fig.\ref{figbulk}.
The effect of including the $\mathcal{O}(T^8)$ in $c_s^2$
effectively produces a peak around  $T_c$, not present to
$\mathcal{O}(T^6)$ \cite{Fernandez-Fraile:07}. The speed of sound
is not the only relevant effect yielding a sizable peak:
unitarization of the cross section entering
$\mathnormal{\Gamma}_p$
\cite{Fernandez-Fraile:05} is also crucial to $\mathcal{O}(T^8)$. Considering unitarized partial waves for
$\pi\pi$ scattering (ChPT is only perturbatively unitary) improves
the high energy behavior (and therefore the high temperature one)
  and generates dynamically the $f_0(600)$ and $\rho(770)$ resonance
poles. Consistently,  we have chosen the
 values of the low-energy constants $\bar l_i$  entering
pion scattering (they can be found in \cite{Fernandez-Fraile:07-2})
so that the mass and width of the $\rho$ are at their physical
values for $T=0$. As we discuss below, the $\bar l_i$ dependence is
crucial in the present analysis. In the chiral limit, the transition
peak is almost unchanged and so is $T_c$, unlike $T_c^\chi$,  which
indicates that chiral restoration is not the main
 source of this effect. Our massless results are in reasonable agreement with
 a recent kinetic theory analysis \cite{Chen:07}.  We also obtain a low-$T$ peak, which disappears in the chiral limit.
In our regime  and for $T\ll m_\pi$, $n_B(E_p)\simeq e^{-m_\pi/T}e^{-p^2/2m_\pi T}$ so
that
 three-momenta  $p=\mathcal{O}(\sqrt{m_\pi T})$ and taking the leading order for $\mathnormal{\Gamma}_p$
 \cite{Fernandez-Fraile:05} and $c_s^2\simeq T/m_\pi+\dots$ \cite{Gerber:89},
 eq.(\ref{bulk}) becomes:
 \begin{equation}\label{bulklowT}
 \zeta (T)\simeq 13.3 \frac{f_\pi^4 \sqrt{T}}{m_\pi^{3/2}} \qquad
\mbox{for} \quad T\ll m_\pi,
 \end{equation}
where $f_\pi$ is the pion decay constant. The above behavior  is
consistent with  nonrelativistic  kinetic theory
\cite{Gavin:1985ph} where $\zeta$ and $\eta$ are expected to be comparable at low $T$. Thus, $\zeta(T)$ increases for very low $T$ and
has to decrease at some point to match the asymptotic vanishing
behavior, thus explaining the low-$T$ maximum.

\begin{figure}
\centerline{\includegraphics[width=6.5cm]{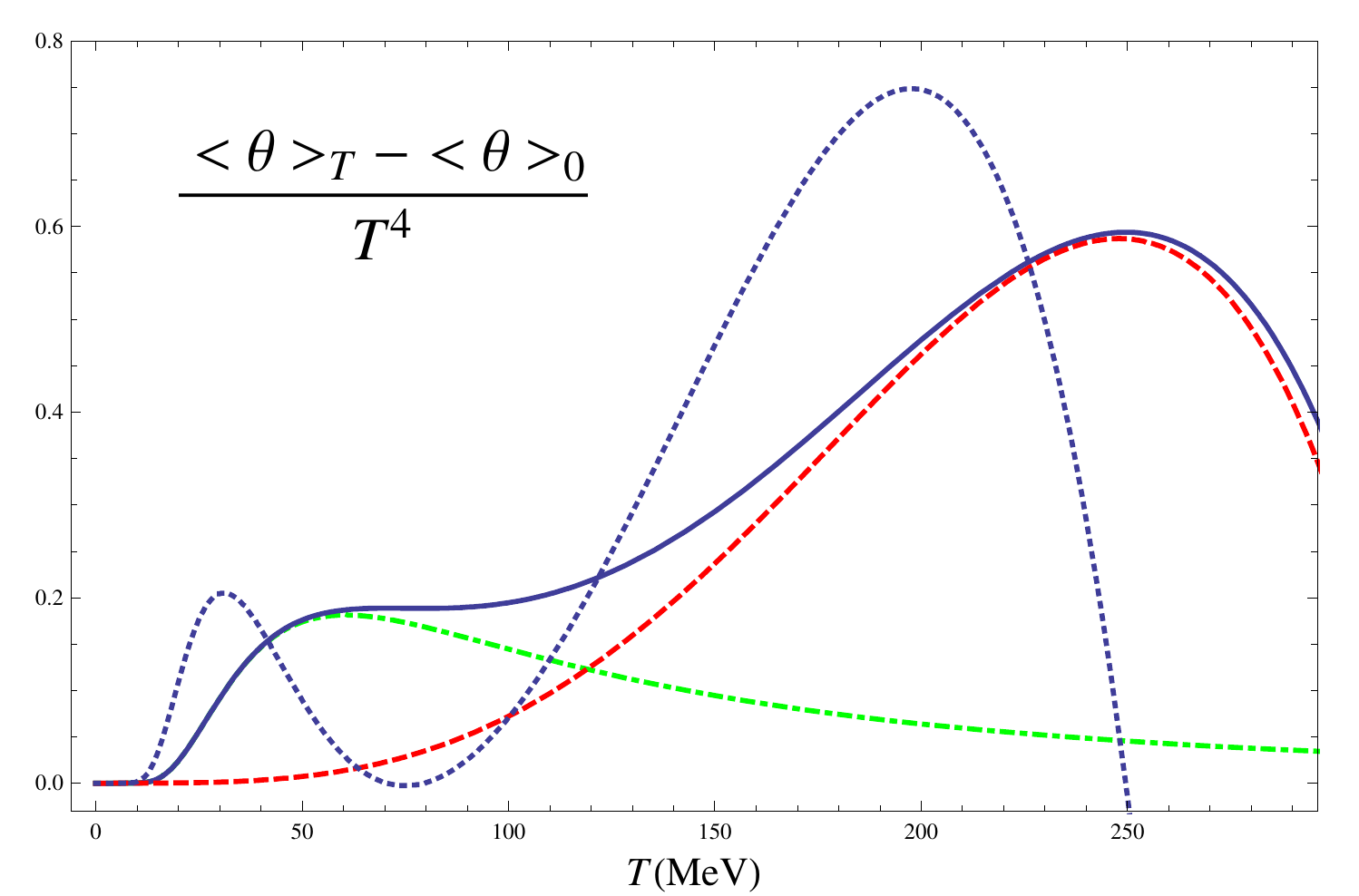}}
\caption{Thermal expectation value of the trace anomaly for the pion
gas. The dashed-dotted green and continuous blue lines are,
respectively, the $\mathcal{O}(T^6)$ and $\mathcal{O}(T^8)$ results.
The dashed red line corresponds to the $\mathcal{O}(T^8)$ result for
massless pions (the $\mathcal{O}(T^6)$ order  vanishes for
$m_\pi=0$). The dotted blue line is $T\frac{\partial}{\partial
    T}\frac{\langle\theta\rangle_T-\langle\theta\rangle_0}{T^4}$.
}\label{figtraceanomaly1}
\end{figure}

Let us now evaluate conformal-breaking contributions for the pion
gas. First, it is instructive to recall the QCD  result for the
trace anomaly \cite{Collins:1976yq}:
\begin{equation} \left(T^\mu_\mu\right)_{QCD}=\frac{\beta(g)}{2g}G_{\mu\nu}^a
G^{\mu\nu}_a+(1+\gamma_m(g))\bar{q}M q, \label{qcdanom}
\end{equation}
where the renormalization group functions are, perturbatively,
$\beta(g)=\mathcal{O}(g^3)$,
$\gamma_m(g)= \mathcal{O}(g^2)$. The first term is the conformal
anomaly proportional to the gluon condensate. The second one
comes from the explicit breaking in the QCD lagrangian, $M$ being
the quark mass matrix.
 For the pion gas, using  the thermodynamic identity
\begin{equation}
\langle\theta\rangle_T=T^5\frac{\mathrm{d}}{\mathrm{d}T}\left(\frac{P}{T^4}\right), \label{therid}
\end{equation}
we represent in Fig.\ref{figtraceanomaly1} the trace anomaly
 to different orders in the pressure, as well as the $T$-function
 appearing in the r.h.s. of (\ref{kharzeev}). We observe clearly  the same two-peak
 structure as the bulk viscosity, with similar features.

 The low-$T$ peak
 disappears in the chiral limit.
  Its contribution comes then from explicit conformal breaking.  Calculating only
 the first nonvanishing  order in ChPT, either using (\ref{therid}) or
 evaluating directly the energy-momentum correlators, we get:
\begin{eqnarray}
    \langle\theta\rangle_T-\langle\theta\rangle_0&=&3m_\pi^2g_1(m_\pi,T)+\mathcal{O}(f_\pi^{-2})
    \nonumber\\&=&
    2m_q\left(\langle
    \bar q q\rangle_T-\langle
    \bar q q\rangle_0\right)+ \mathcal{O}(f_\pi^{-2}),
\label{anomalychpt}
 \end{eqnarray}
where $m_q=m_u=m_d$ and we  formally account for different chiral
orders by their $f_\pi$ power. The function $g_1$ is the thermal
correction to the free pion propagator $G(x=0)$ \cite{Gerber:89}.
Comparing with the QCD expression (\ref{qcdanom}) the factor of two
in (\ref{anomalychpt}) for the quark condensate is perfectly
consistent with the result \cite{Agasian:2001sv} showing that the
quark and gluon contributions to the trace anomaly are identical at
low temperatures. Now, $g_1(T)/T^4$ has a maximum at $T\simeq
2m_\pi/5\simeq 60$ MeV, which is the low-$T$ peak in
Fig.\ref{figtraceanomaly1} and the source for the first peak of the
bulk viscosity.

The transition peak only shows up at $\mathcal{O}(T^8)$ and
survives in the chiral limit, where its origin is purely
anomalous. It comes from ChPT interactions involving dimensionful
couplings, like $f_\pi$, and is therefore suppressed at low
 temperatures \cite{Leutwyler:92}. For massive
pions, the value of the peak and its position are almost unchanged
with respect to the chiral limit,  the difference being even smaller
than the simple extrapolation of the quark condensate contribution
in (\ref{anomalychpt}) with $\langle\bar q q\rangle_T$  to
$\mathcal{O}(T^8)$, which represents around a 10\% correction in the
critical region. The fermion contribution is also subdominant in
lattice analyses \cite{Cheng:08}. These results show again that the
nature of this effect is not likely to be related to chiral symmetry
restoration but rather to other QCD critical effects like
deconfinement. The correlation with the bulk viscosity is again clear.
In fact, in the chiral limit the function between brackets in
(\ref{kharzeev}) and $15(c_s^2-1/3)^2=\zeta/\eta$ have their maximum
 at the same $T_c=e^{-5/8}\mathnormal{\Lambda}_p$ with $\mathnormal{\Lambda}_p$ given
in \cite{Gerber:89} in terms of $\bar l_1+4\bar l_2$. We recall that
in order to establish the possible correlations between the
conformal anomaly and the bulk viscosity, we have used the same set
of $\bar l_i$ in both figures. For those unitarized values,
$T_c\simeq 220$ MeV.  Using perturbative values, for instance those
given in \cite{Gerber:89} fixed to reproduce pion scattering
lengths, the critical peak is about three times smaller and
$T_c\simeq 148$ MeV,  while $T_c^\chi$ varies only about 10 MeV from
one set to another. We get exactly the same drastic reduction of the
critical peak and shift of $T_c$ in the bulk viscosity. The presence
of resonances is then crucial to yield a sizable effect in the
transition peak, whose dominant contribution comes from the gluon
condensate.

Regarding the  $\omega_0(T)$ function defined through
(\ref{kharzeev}), in the chiral limit it grows  linearly with $T$,
reaching $\omega_0\sim 400$ MeV at the transition. In
the massive case, taking $\vert\epsilon_0\vert=f_\pi^2m_\pi^2$, the
ChPT lowest order, we get $\omega_0(T_c)\sim$ 1 GeV,  almost
constant from $T\sim 150$ MeV onwards. These values are in
reasonable agreement with the estimates in \cite{Karsch:07}. On the
other hand, from (\ref{bulklowT}) we get $\omega_0(T)\simeq 0.13
m_\pi^{7/2}/(f_\pi^2 \sqrt{T})$ for $T\rightarrow 0^+$.

The numerical values of the trace anomaly in
Fig.\ref{figtraceanomaly1} are not far from the lattice values
\cite{Cheng:08} for low $T$, but they are about a factor
of 10 smaller near $T_c$. The increasing of degrees of
freedom due to heavier states, not included in our approach, is
clearly important  in that region. For instance, the
$\mathcal{O}(T^8)$ pressure in the chiral limit is proportional to
$N_f^2(N_f^2-1)$ \cite{Leutwyler:92} so that changing from two to
three flavors, which are not Boltzmann suppressed near the
transition, increases significatively the anomaly. In addition,
using a simple free Hadron Resonance Gas approach
\cite{Karsch:07}, the $\pi$,$\rho$,$\sigma$ contributions to the
anomaly amount only to a 5\% of all baryon and meson states up to
2.5 GeV. In fact, although we get $\zeta/s\simeq 0.02$ at the
transition peak, still much smaller than $\eta/s\simeq 0.25$, we
would get a larger value if we assume that the introduction of
heavier states increase the anomaly, and that implies an increase
of the transition strength and a strong reduction of $c_s^2$
\cite{Karsch:07}. As an indication, setting $c_s^2=0$ in
(\ref{bulk}) we get $\zeta/s\sim 1$ at $T_c$.

 We have seen  that it is crucial to include correctly the
effect of the $\rho$ resonance. On the other hand, the
$f_0(600)/\sigma$  is expected to be related to chiral restoration.
Regarding bulk viscosity, it has been suggested in
\cite{Paech:2006st}, within mean field theory, that any dynamic
scalar field $\sigma$ should contribute to
$\zeta\propto\mathnormal{\Gamma}_\sigma/m_\sigma^2$, which may be large  near the
critical region by mass reduction, for instance in the Linear Sigma
Model (LSM) context. Within unitarized ChPT, the dynamically
generated $f_0(600)$ pole
 undergoes a significant mass reduction towards $2m_\pi$ governed
by chiral restoration, remaining a broad state with sizable width
near the transition \cite{Fernandez-Fraile:07-2}. Interestingly,
from \cite{Fernandez-Fraile:07-2}, we find that
$\mathnormal{\Gamma}_\sigma/m_\sigma^2$ has a peak at $T\sim$ 180 MeV, where the
pole mass reaches threshold. For higher $T$  the width still
decreases (by phase space reduction) while the mass remains close to
threshold. This critical value is very close to the one obtained in
\cite{Paech:2006st} for the LSM assuming a $T$-independent width.
 However, as discussed above, these chiral
restoration effects are  likely to be subdominant.

The large-$N_c$ limit  is also revealing. The counting of the
$\bar l_i$ can be extracted from the $L_i$ ($N_f=3$)
\cite{Gasser:1984gg} while $f_\pi^2=\mathcal{O}(N_c)$.  We get
$\mathnormal{\Gamma}_p\sim\mathcal{O}(N_c^{-2})$ and, in the
chiral limit,
$\langle\theta\rangle_T\sim\mathcal{O}(N_c^{-1})\sim(c_s^2-1/3)$
so that $\zeta\sim\mathcal{O}(1)$ and
$\zeta/\eta\sim\mathcal{O}(N_c^{-2})$, parametrically suppressed
as expected. Now, taking into account the critical behavior,
$T_c\sim \mathcal{O}(e^{N_c})$ and
$\langle\theta\rangle_{T_c}\sim\mathcal{O}(e^{N_c}/N_c^2)$. This
 large dependence is another indication of the dominance of confinement
 over chiral restoration, comparing with  the chiral
$T_c^\chi=\mathcal{O}(N_c)$. Also, $\langle\theta\rangle\propto
L_3$, which in  large-$N_c$  includes a term proportional to
the gluon condensate \cite{Espriu:1989ff}.  Comparing with the QCD
expressions in \cite{Arnold:2006fz}, we agree except for the overall
$\mathcal{O}(N_c^2)$ constants in the pressure which  count
 the degrees of freedom. For massive pions, the
above chiral limit scaling is only reached asymptotically for
large $T$, while for any $T$ we get
$\zeta/\eta\sim\mathcal{O}(1)\sim
\langle\theta\rangle_T-\langle\theta\rangle_0$ with
$\zeta\sim\mathcal{O}(N_c^2)$, compatible with (\ref{bulklowT}).

Summarizing, we have shown, within unitarized ChPT, that the
massive pion gas develops a strong correlation between  bulk
viscosity and the conformal anomaly. Both quantities show a
low-temperature peak coming from mass conformal breaking and another
one at the critical temperature remaining in the chiral limit
and mainly dominated by  gluon condensate contributions not
related to chiral restoration. The dynamically generated light
resonances are essential to obtain sizable effects at the
transition. Different estimates indicate that heavier states could
yield a larger bulk viscosity near the transition,  leading
to observable effects in  heavy ion collisions.

\begin{acknowledgments} We are grateful to D.Kharzeev, G.D. Moore and C. Pica
for very useful comments. Research partially funded by  research
contracts FPA2004-02602, FPA2005-02327, FPA2007-29115-E, PR34-1856-BSCH, UCM-CAM
910309
 and
 FPI-BES-2005-6726.
\end{acknowledgments}

\end{document}